\documentclass[prd, reprint,superscriptaddress,preprintnumbers, nolongbibliography,
nofootinbib,
 amsmath,amssymb,
 aps]{revtex4-2}
\usepackage{multirow}
\usepackage{upgreek}
\usepackage{tensor}
\usepackage{float}
\usepackage{latexsym}
\usepackage{graphicx}
\usepackage{amssymb}
\usepackage{amsmath}
\usepackage{amsfonts}
\usepackage{bm}
\usepackage[hidelinks]{hyperref}
\usepackage{color}
\usepackage{cool}
\usepackage{booktabs}
\usepackage{dcolumn}
\usepackage{slashed}
\usepackage{tabularx}
\usepackage{multirow}
\usepackage{mathrsfs}

\begin{document}

\title{Constraints on interacting dark energy revisited: implications for the Hubble tension}


\author{Gabriel A. Hoerning}\email{gabrielamancio.hoerning@postgrad.manchester.ac.uk}
\thanks{these two authors contributed equally.}
\affiliation{Jodrell Bank Centre for Astrophysics, Department of Physics \& Astronomy,\\
The University of Manchester, Oxford Road, Manchester, M13 9PL, U.K.}
\affiliation{Instituto de F\'isica, Universidade de S\~ao Paulo \\
Rua do Mat\~ao, 1371, Butant\~a, CEP 05508-090, S\~ao Paulo, SP, Brazil}

\author{Ricardo G. Landim}\email{ricardo.landim@port.ac.uk}
\thanks{these two authors contributed equally.}
\affiliation{Institute of Cosmology and Gravitation, University of Portsmouth\\ Dennis Sciama Building, Portsmouth PO1 3FX, United Kingdom}
\affiliation{ Technische Universit\"at M\"unchen, Physik-Department T70, \\James-Franck-Strasse 1, 85748, Garching, Germany\label{inst:tum}}

\author{Luiza O. Ponte}
\affiliation{Instituto de F\'isica, Universidade de S\~ao Paulo \\
Rua do Mat\~ao, 1371, Butant\~a, CEP 05508-090, S\~ao Paulo, SP, Brazil}

\author{Raphael P. Rolim}
\affiliation{Instituto de F\'isica, Universidade de S\~ao Paulo \\
Rua do Mat\~ao, 1371, Butant\~a, CEP 05508-090, S\~ao Paulo, SP, Brazil}

\author{Filipe B. Abdalla}
\affiliation{School of Astronomy and Space Science, University of Science and Technology of China, \\
Hefei, Anhui 230026, China}
\affiliation{Department of Physics and Electronics, Rhodes University, \\PO Box 94, Grahamstown, 6140, South Africa}

\author{Elcio Abdalla}
\affiliation{Instituto de F\'isica, Universidade de S\~ao Paulo \\
Rua do Mat\~ao, 1371, Butant\~a, CEP 05508-090, S\~ao Paulo, SP, Brazil}
\affiliation{Universidade Estadual da Paraíba, Rua Baraúnas, 351, Bairro Universitário, Campina Grande, Brazil}
\affiliation{Departamento de Física, Centro de Ciências Exatas e da Natureza, Universidade Federal da Paraíba, CEP 58059-970, João Pessoa, Brazil}

\date{\today}

\begin{abstract}
We revisit a class of coupled dark energy models where dark energy interacts
with dark matter via phenomenological energy exchange terms. We include the
perturbation of the Hubble rate in the interaction term, which was absent in
previous works. We also consider more recent data sets such as cosmic microwave
background (CMB) anisotropies from \text{Planck} 2018, type I-a supernovae (SNIa)
measurements from Pantheon+ and data from baryon acoustic oscillations (BAO),
and redshift space distortions (RSD). One of the models presents a strong
incompatibility when different cosmological datasets are used. We analyze the
influence of the SH0ES Cepheid host distances on the results and, although for
one model the discrepancy of $H_0$ is reduced to $3.2\sigma$ when compared to the value obtained by the \textit{Planck} collaboration and
$3.6\sigma$ when compared to the SH0ES team, joint analysis is incompatible. Including
BAO with RSD shows incompatibility with SH0ES for all models considered here. We
perform a model comparison and although there is a preference for interacting dark energy over $\Lambda$CDM for all the models for joint analysis
CMB+BAO+RSD+SNIa, most of the 2D contours do not overlap. We conclude that the models of interactions in the dark
sector considered in this paper are not flexible enough to fit all the
cosmological data including values of $H_0$ from SH0ES in a statistically
acceptable way.  Therefore, the addition of one extra degree of freedom (i.e. the coupling between dark matter and dark matter) does not help enough to alleviate the already existing tension in the vanilla $\Lambda$CDM, suggesting that the models would need to be
modified to include further flexibility of predictions to help elucidate this issue.

  \end{abstract}

\maketitle

\section{Introduction}


The lack of a satisfactory theoretical explanation for the nature of the cosmological constant, used as the standard candidate for the late time accelerated expansion of the Universe \cite{perlmutter1999,*reiss1998}, added to the existing tensions in the $\Lambda$-Cold Dark Matter ($\Lambda$CDM) model, such as the Hubble tension \cite{Aghanim:2018eyx,Riess:2021jrx,DiValentino:2021izs, Dainotti:2021pqg,*Dainotti:2022bzg,*Lenart:2022nip,*Bargiacchi:2023jse, *Dainotti:2023ebr,*Bargiacchi:2023rfd}, make room for alternative explanations of Dark Energy (DE). There has been a plethora of alternative candidates to explain the cosmic acceleration, such as scalar and vector fields \cite{peebles1988,*ratra1988,*Frieman1992,*Frieman1995,*Caldwell:1997ii,*Padmanabhan:2002cp,*Bagla:2002yn,*ArmendarizPicon:2000dh,*Brax1999,*Copeland2000,*Vagnozzi:2018jhn,*Koivisto:2008xf,*Bamba:2008ja,*Emelyanov:2011ze,*Emelyanov:2011wn,*Emelyanov:2011kn,*Kouwn:2015cdw,*Landim:2015upa,*Landim:2016dxh,*Banerjee:2020xcn}, metastable DE \cite{Szydlowski:2017wlv,*Stachowski:2016zpq,*Stojkovic:2007dw,*Greenwood:2008qp,*Abdalla:2012ug,*Shafieloo:2016bpk,*Landim:2016isc, *Landim:2017kyz,*Landim:2017lyq}, holographic DE \cite{Hsu:2004ri,*Li:2004rb,*Pavon:2005yx,*Wang:2005jx,*Wang:2005pk,*Wang:2005ph,*Wang:2007ak,*Landim:2015hqa,*Li:2009bn,*Li:2009zs,*Li:2011sd,*Saridakis:2017rdo,*Mamon:2017crm,*Mukherjee:2016lor,*Feng:2016djj,*Herrera:2016uci,*Forte:2016ben,*Landim:2022jgr}, models using extra dimensions \cite{dvali2000}, alternative fluids \cite{Landim:2021www, *Landim:2021ial, Adil:2023jtu, *Malekjani:2023dky, *Colgain:2022rxy, *Gangopadhyay:2023nli, *Gangopadhyay:2022bsh}, etc. Among the many possibilities, interacting DE (IDE) \cite{Wetterich:1994bg,*Amendola:1999er,*Guo:2004vg,*Cai:2004dk,*Guo:2004xx,*Bi:2004ns,*Gumjudpai:2005ry,*Yin:2007vq,*Abdalla:2014cla,*Costa:2014pba,*Landim:2015poa,*Landim:2015uda,*Marcondes:2016reb,*Landim:2016gpz,Wang:2016lxa,Farrar:2003uw,*micheletti2009,*Yang:2017yme,*Marttens:2016cba,*Yang:2017zjs,*Costa:2018aoy,*Yang:2018euj,*Landim:2019lvl,*Vagnozzi:2019kvw,*Johnson:2021wou, *Johnson:2020gzn,Costa:2013sva,Costa:2016tpb} can help alleviating the coincidence problem \cite{Olivares:2005tb} and the Hubble tension \cite{DiValentino:2019ffd,*DiValentino:2019jae,Lucca:2020zjb}.

The interaction between Dark Matter (DM) and DE has been vastly explored in the literature, using different forms for the interaction. While the main aim is to have a canonical Field theory description of the Dark Sector, it is still far from reach. A full description should include gravity, thus non-renormalizability is a severe constraint and possibly has to go through the construction of quantum gravity \cite{Cicoli:2023opf} (see also \cite{Abdalla:2020ypg}).

A sub-class of IDE consists of an interaction at the background level proportional to a weighted sum of the energy densities of DM and DE ($Q=H(\lambda_1 \rho_{\rm dm}+\lambda_2 \rho_{\rm de})$ -- see \cite{Wang:2016lxa} for a review). Latest constraints on the coupling constants were presented in \cite{Costa:2016tpb}, using \textit{Planck} 2015, while forecasts on such models have been produced for several upcoming observational programs \cite{Costa:2019uvk, *Bachega:2019fki,* Costa:2021jsk,*Xiao:2021nmk}.

Although such kernels for IDE have been widely investigated, none of the previous works took into account the perturbation of the Hubble rate in the perturbation equations.  As already pointed out in \cite{Valiviita:2008iv}, the unperturbed $H$ appeared in the interaction
in many previous works to simplify the solution of the equations,  not
resulting from physical principles. Authors of \cite{Gavela:2010tm} decided that $H$ is not the
global (average) expansion rate but a local expansion rate and hence perturbed $H$. This was also required by gauge invariance, as pointed out in \cite{Gavela:2010tm} where the authors identified the neglected term from the Hubble rate and corrected the perturbation equations. However, such a correction has not been done yet for the interaction considered here. The presence of the perturbation of the Hubble rate is required by gauge invariance although one may simply guess that there is no physical reason to ignore such a term since the perturbation of the product of two variables acts as a chain rule.

In addition to the correction of the perturbation equations, we use more recent cosmological data sets than the ones assumed in \cite{Costa:2016tpb} to constrain the models, including cosmic microwave background (CMB) anisotropy measurements from \textit{Planck} 2018 \cite{Planck:2018nkj}, type I-a supernovae (SNIa) from Pantheon+ \cite{Scolnic:2021amr}, data from baryon acoustic oscillations (BAO) and redshift space distortions (RSD). Additionally, we performed some analysis using the SH0ES Cepheid host distance anchors in the likelihood \cite[R22;][]{Riess:2021jrx}, aiming to investigate whether or not these models can alleviate the Hubble tension. The tension arises from the $5\sigma$ discrepancy between the value of $H_0$ obtained by \textit{Planck} ($67.4\pm 0.5$ km/s/Mpc) \cite{Aghanim:2018eyx} and that derived from the SH0ES team ($73.04\pm 1.04$ km/s/Mpc). 

After a detailed analysis of the different kernels, we obtained more stringent constraints on these IDE models,  however, for any model the 2D contours do not overlap at $1\sigma$ when the combination of datasets CMB, BAO, RSD, and SNIa is taken into account. The calculated evidences indicate a preference for IDE models over $\Lambda$CDM across all individual and partial combinations of datasets. However, when all datasets are combined, the preference shifts slightly in favour of $\Lambda$CDM. For one of the models, the value obtained for $H_0$ ($69.18^{+0.26}_{-0.27}$ km/s/Mpc) differs by $3.2\sigma$ from both the Planck collaboration measurements and $3.6\sigma$ for the R22 measurements in the $\Lambda$CDM model.

This paper is organized in the following manner. Sec. \ref{sec:idedm} presents the models and the background and perturbation equations. In Sec. \ref{sec:results} we show the cosmological datasets used and the results. Sec. \ref{sec:conclusions} is reserved for conclusions. We use Natural units ($c=\hbar=1$) throughout the text.

\section{Interacting dark energy models}\label{sec:idedm}
Interaction in the dark sector can be a complex and difficult issue. We assume here that both DE and DM are described by ideal fluids -- thus a non-Lagrangian description. In that case, with an interaction between DE and DM  the energy-momentum tensor of each component is no longer individually conserved, but both tensors satisfy
\begin{equation}
    \nabla_\mu T^{\mu \nu}_{(i)} = Q^\nu_{(i)}\,,
\end{equation}
where the index $i$ represents either the DM component (with index `dm') or the DE component (with index `de'). The four-vector $Q^{\nu}_{(i)}$ regulates the flux between the two species of the dark sector and due to the Bianchi identities, the total energy-momentum tensor should be conserved, implying that $Q^{\nu}_{\rm dm}=-Q^{\nu}_{\rm de}$. Many different phenomenological couplings $Q^{\nu}$ have been used in the literature. Here we restrict to the class of models where the interaction is present only in the zeroth component of the coupling vector, i.e. $Q^i_{\rm dm,de}=0$.\footnote{In these models the intrinsic momentum transfer potential $f_A$ (see Sec. II of \cite{Valiviita:2008iv}) is proportional to the momentum transfer $f_A = Q_A \theta_A/k^2$.}

Assuming a Friedmann-Lemaitre-Robertson-Walker metric, the continuity equations for DE and DM are
\begin{align}
    \dot{\rho}_{\rm dm}+3\mathcal{H}\rho_{\rm dm}&=a^2 Q^0_{\rm dm}= a Q\,,\\
     \dot{\rho}_{\rm de}+3\mathcal{H}(1+w)\rho_{\rm de}&=a^2 Q^0_{\rm de}= -a Q\,,
\end{align}
where $w$ is a constant DE equation of state, a dot represents a conformal time derivative, $\mathcal{H}=aH$ is the Hubble rate for the conformal time and we assume the interaction to be given by $Q=H(\lambda_1\rho_{\rm dm}+\lambda_2\rho_{\rm de})$ \cite{He:2008si}. A positive $Q$ corresponds to DE being transformed into DM (as in the case of the alleviation of the coincidence problem in \cite{Olivares:2005tb}), while negative $Q$ means
the transformation in the opposite direction.

Considering the usual three different choices for the coupling constants ($\{\lambda_1\neq 0$, $\lambda_2=0\}$, $\{\lambda_1= 0$, $\lambda_2\neq0\}$, $\{\lambda_1=\lambda_2 \equiv \lambda\}$) it is possible to find analytic solutions for the continuity equations above. For $\lambda_1\neq 0$, $\lambda_2=0$ we have 
\begin{align}
    \rho_{\rm dm}& =  \rho_{ \rm dm,0} a^{-3(1+w^{\rm eff}_1)}\,,\\
    \rho_{\rm de} & = \rho_{\rm de,0}a^{-3(1+w)}+\lambda_1\frac{\rho_{\rm dm,0}a^{-3(1+w)}}{3(w-w^{\rm eff}_1)}\bigg[1-a^{3(w-w^{\rm eff}_1)}\bigg]\,,
\end{align}
where $w^{\rm eff}_1 = -\lambda_1/3$. 

For $\lambda_1= 0$, $\lambda_2\neq0$ the solutions are \cite{Lucca:2020zjb}
\begin{align}\label{eq:lucca_dm}
    \rho_{\rm dm}& =  \rho_{\rm dm,0}a^{-3}+\lambda_2\frac{\rho_{\rm de,0}a^{-3}}{3w^{\rm eff}_2}\bigg[1-a^{-3w^{\rm eff}_2}\bigg]\,,\\
    \rho_{\rm de} &=  \rho_{ \rm de,0} a^{-3(1+w^{\rm eff}_2)}\,,\label{eq:lucca_de}
\end{align}
where $w^{\rm eff}_2=w+\lambda_2/3$.

The solution for the case $\lambda_1=\lambda_2\equiv \lambda$ is \cite{Olivares:2005tb}
\begin{widetext}
\begin{align}
     \rho_{\rm dm}& = w_{\rm eff}^{-1} \bigg\{\bigg[\bigg(1+w+\frac{\lambda}{3}\bigg)\rho_{\rm dm,0}+\frac{\lambda}{3}\rho_{\rm de,0}\bigg](a^{S_-} - a^{S_+})+\rho_{\rm dm,0}(S_- a^{S_-}- S_+a^{S_+})\bigg\}\,,\\
    \rho_{\rm de} &=  w_{\rm eff}^{-1} \bigg\{\bigg[\frac{\lambda}{3}\rho_{\rm dm,0}-\bigg(1-\frac{\lambda}{3}\bigg)\rho_{\rm de,0}\bigg](a^{S_+} - a^{S_-})+\rho_{\rm de,0}(S_- a^{S_-}- S_+a^{S_+})\bigg\}\,,
\end{align}
\end{widetext}
where $w_{\rm eff}=(w^2+4\lambda w/3)^{1/2}$ and $S_{\pm}=-(1+w/2)\mp w_{\rm eff}/2$.

An advantage of the IDE models is that the coincidence problem can be alleviated. When the interaction is either proportional to $\rho_{\rm dm}$ or $\rho_{\rm dm}+\rho_{\rm de}$, the ratio $\rho_{\rm dm}/\rho_{\rm de}$ reaches a plateau for very high or very low redshifts. Therefore, for these models, the ratio is practically constant. One can then use the continuity equations for DM and DE  to obtain the solutions of the equation  $\dot{r}=0$, where $r\equiv\rho_{\rm dm}/\rho_{\rm de}$. The two solutions are \cite{He:2008si} 
\begin{widetext}
\begin{align}
 \lambda_1 r_+ & = -\frac{3}{2}\bigg(w+\frac{\lambda_1}{3}+ \frac{\lambda_2}{3}\bigg)+\frac{3}{2}  \sqrt{w^2+\frac{2}{3}w(\lambda_1+\lambda_2) +\frac{1}{9}(\lambda_1-\lambda_2)^2}\,,\label{eq:rplus}\\
 \lambda_1 r_- & = -\frac{3}{2}\bigg(w+\frac{\lambda_1}{3}+ \frac{\lambda_2}{3}\bigg)-\frac{3}{2}  \sqrt{w^2+\frac{2}{3}w(\lambda_1+\lambda_2) +\frac{1}{9}(\lambda_1-\lambda_2)^2}\,.\label{eq:rminus}
\end{align}
\end{widetext}
These equations will be useful later.

The linear order perturbation equations for DM and DE can be found using the gauge-invariant equations presented in \cite{Ma:1995ey,Valiviita:2008iv,Gavela:2010tm}. In the synchronous gauge, they are
\begin{widetext}
\begin{align}
    \dot{\delta}_{\rm dm} =& -\theta_{\rm dm}-\frac{\dot{h}}{2}+\mathcal{H}\lambda_2\frac{\rho_{\rm de}}{\rho_{\rm dm}}(\delta_{\rm de}-\delta_{\rm dm})+\bigg(\lambda_1+\lambda_2\frac{\rho_{\rm de}}{\rho_{\rm dm}}\bigg)\bigg(\frac{kv_T}{3}+\frac{\dot{h}}{6}\bigg)\,,\label{eq:delta_dm}\\
    \dot{\theta}_{\rm dm} = & -\mathcal{H}\theta_{\rm dm}-\bigg(\lambda_1+\lambda_2\frac{\rho_{\rm de}}{\rho_{\rm dm}}\bigg)\mathcal{H}\theta_{\rm dm}\,,\\
    \dot{\delta}_{\rm de} =& -(1+w)\bigg(\theta_{\rm de}+\frac{\dot{h}}{2}\bigg)-3\mathcal{H}(1-w)\delta_{\rm de}+\mathcal{H}\lambda_1\frac{\rho_{\rm dm}}{\rho_{\rm de}}(\delta_{\rm de}-\delta_{\rm dm})\nonumber\\&-3\mathcal{H}(1-w)\bigg[3(1+w)+\lambda_1\frac{\rho_{\rm dm}}{\rho_{\rm de}}+\lambda_2\bigg]\frac{\mathcal{H}\theta_{\rm de}}{k^2}-\bigg(\lambda_1\frac{\rho_{\rm dm}}{\rho_{\rm de}}+\lambda_2\bigg)\bigg(\frac{kv_T}{3}+\frac{\dot{h}}{6}\bigg)\,,\label{eq:delta_de}\\
    \dot{\theta}_{\rm de} = & 2\mathcal{H}\theta_{\rm de}\bigg[1+\frac{1}{1+w}\bigg(\lambda_1\frac{\rho_{\rm dm}}{\rho_{\rm de}}+\lambda_2\bigg)\bigg]+\frac{k^2}{1+w}\delta_{\rm de}\,,
\end{align}
\end{widetext}
where the DE effective sound speed
is set to one, the adiabatic sound speed is set to be $w$ and the centre of mass velocity for the total fluid $v_T$ is defined as \cite{Gavela:2010tm} 
\begin{equation}\label{eq:vt}
   (1+w_T) v_T=\sum_a (1+w_a)\Omega_a v_a\,.
\end{equation}

There is an extra term in the equations for the density contrasts (\ref{eq:delta_dm}) and (\ref{eq:delta_de}) due to the perturbation of the Hubble rate $\delta H$, which was absent in all previous works in the literature that considered the assumed phenomenological interaction (see e.g. \cite{He:2008si, Costa:2016tpb}). This extra term is:
\begin{equation}
    \delta H = \frac{kv_T}{3}+\frac{\dot{h}}{6}\,.
\end{equation}

The adiabatic initial conditions for DM and DE in the synchronous gauge are given by \cite{He:2008si,costa2014observational}
\begin{align}
    \delta_{\rm de}^{(i)}&=\frac{3}{4}\delta_r^{(i)} \bigg(1+w+\frac{\lambda_1}{3}r+\frac{\lambda_2}{3}\bigg)\,,\label{eq:delta_ini_de}\\
    \delta_{\rm dm}^{(i)}&=\frac{3}{4}\delta_r^{(i)} \bigg(1-\frac{\lambda_1}{3}-\frac{\lambda_2}{3}\frac{1}{r}\bigg)\,,\label{eq:delta_ini_dm}\\
    v_{\rm de}^{(i)} &= v_r^{(i)}\,,
    \end{align}
where the index `r' represents radiation, while for the other species, they remain the same as in $\Lambda$CDM. The residual freedom of the synchronous gauge is fixed by choosing a comoving frame such that the DM velocity is zero.   These initial conditions can be simplified by noticing that in the early Universe either the ratio $\rho_{\rm dm}/\rho_{\rm de}$ is constant (for interactions $\propto \rho_{\rm dm}$ or $\propto \rho_{\rm dm}+\rho_{\rm de}$) or $\rho_{\rm de}/\rho_{\rm dm}\approx 0$ (for $Q\propto \rho_{\rm de}$) \cite{He:2008si,Olivares:2005tb}. Using Eqs. (\ref{eq:rplus}) and (\ref{eq:rminus}) the initial conditions become, for $Q \propto \rho_{\rm dm}$
\begin{align}
    \delta_{\rm de}^{(i)}= \delta_{\rm dm}^{(i)}=\frac{3}{4}\delta_r^{(i)} \bigg(1-\frac{\lambda_1}{3}\bigg)\,,
    \end{align}
while for $Q\propto \rho_{\rm de}$ we have
\begin{align}
    \delta_{\rm de}^{(i)}&= \frac{3}{4}\delta_r^{(i)} \bigg(1+w+\frac{\lambda_2}{3}\bigg)\,,\\
    \delta_{\rm dm}^{(i)}&= \frac{3}{4}\delta_r^{(i)}\,.
    \end{align}
    The initial conditions for $Q\propto \rho_{\rm dm} + \rho_{\rm de}$ remain the ones in Eqs. (\ref{eq:delta_ini_de}) and (\ref{eq:delta_ini_dm}) with $r$ given by Eq. (\ref{eq:rplus}) and $\lambda_1=\lambda_2$. 
    
    To avoid complex or negative energy densities and early or late time instabilities, the couplings and the equation of state for DE should be restricted to the values presented in Table \ref{tab:restrictions} \cite{Gavela:2009cy, He:2008si, Costa:2016tpb}. Although the phantom behaviour might be problematic because it leads to an increasing energy density in some cases, or to an inconsistent relativistic energy-momentum relation, if one uses a scalar field with the opposite sign kinetic term, we leave the equation of state free to reach values smaller than $-1$ for the sake of completeness.

    \begin{table}[h]
        \centering
        \setlength\tabcolsep{3.5pt}
        \renewcommand{\arraystretch}{1.1} 
        \caption{Stability conditions for the models analyzed in the present work.}
        \begin{tabular}{cccc}
        \toprule
        \toprule
            Model & $Q $  & Equation of state & Constraints\\
            \midrule
           I      &  $\lambda_2 H\rho_{\rm de}$   &  $w\neq -1$ &   $\lambda_2<-w$  \\
          II    &  $\lambda_1 H\rho_{\rm dm}$   & $w<-1$ &  $0\leq \lambda_1<-3w$  \\
          III     &  $\lambda H(\rho_{\rm dm} +\rho_{\rm de})$   & $w<-1$ & $0\leq\lambda\leq -3w/4$  \\
           \bottomrule
           \bottomrule
        \end{tabular}
        \label{tab:restrictions}
    \end{table}
    

\subsection{Comparison with model $Q^\nu=\xi H \rho_{\rm de}u_{\rm dm }^\nu$}\label{sec:lucca}

Before moving to the results, we present here the differences between the models we investigated and one well-studied model in the literature (which we will label as Model IV). This scenario is presented in \cite{Lucca:2020zjb} and references therein and we just present the main differences in terms of equations, which will be used to show a comparison in the next section.

The interaction is proportional to the energy density of DE and to the CDM four-velocity $u^\nu_{\rm dm}$. Among other reasons, this parametrization is chosen to avoid momentum transfer in the DM rest frame. At the background level, this model corresponds to our models I or II, but with $w\simeq -1$ ($w=-0.999$ in practice), with the energy density for  DM and DE given by Equations (\ref{eq:lucca_dm}) and (\ref{eq:lucca_de}), respectively.

Because the difference in the interactions only comes from the DM four-velocity, the equations for the DE overdensities are the same, Equations (\ref{eq:delta_dm}) and (\ref{eq:delta_de}), with $\lambda_2=\xi$ and $v_T \sim 0.$\footnote{Because $w\simeq -1$ and $\Omega_{\rm rad}$ was neglected in Eq. (\ref{eq:vt}) \cite{Lucca:2020zjb}.}  Equations for the fluid velocities, on the other hand, become
\begin{align}
   \dot{\theta}_{\rm dm} = & -\mathcal{H}\theta_{\rm dm}\,,\\
    \dot{\theta}_{\rm de} = & 2\mathcal{H}\theta_{\rm de}\bigg[1+\frac{\xi}{1+w}\bigg(1-\frac{\theta_{\rm dm}}{2\theta_{\rm de}}\bigg)\bigg]+\frac{k^2}{1+w}\delta_{\rm de}\,.
\end{align}

\section{Data sets and Results}\label{sec:results}
To constrain the three different IDE models assumed here, we use the most recent data from different surveys:  CMB anisotropies from \textit{Planck} 2018 high-$\ell$
 and low-$\ell$ temperature and polarization power spectra (TT, TE, EE) \cite{Planck:2019nip} and lensing measurements \cite{Planck:2018lbu}, 1701 light curves of SNIa from Pantheon+ \cite{Scolnic:2021amr}, BAO  measurements from 6dFGS \cite{Beutler:2011hx}, MGS \cite{Ross:2014qpa}, BOSS DR12  \cite{BOSS:2016wmc}, DES \cite{abbott2019dark}, eBOSS \cite{alam2021completed}, WiggleZ \cite{blake2011wigglez} and RSD data from 6dFGS \cite{beutler_6df_2012}, Fastsound \cite{okumura2016subaru}, GAMA \cite{blake2013galaxy} and WiggleZ \cite{blake2012wigglez}. We also included in some analysis the SH0ES Cepheid host distance anchors in the likelihood \cite{Riess:2021jrx}, denoting the joint constraints by `$H_0$'. We implemented the background and perturbation equations in a modified version of \texttt{CLASS} \cite{blas2011cosmic,Lucca:2020zjb}\footnote{Available at \url{https://github.com/ricardoclandim/class_IDE_pheno}.} and used the code
 {\tt PLINY} \cite{rollins2015chemical} to execute the Markov Chain Monte Carlo (MCMC) analysis with the Nested Sampling technique \cite{skilling2004nested}. {\tt PLINY} also produces an evidence calculation, though it becomes relevant only when comparing models with some difference between their parametrizations. For this, the Bayes ratio is employed, incorporating the priors of the models into the evaluation. This ratio is defined as:
\begin{equation}\label{bayes}
    \mathcal{B}=\frac{\mathcal{E}(\mathcal{D}|\text{IDE})}{\mathcal{E}(\mathcal{D}|\Lambda\text{CDM})}\, ,
\end{equation}
where $\mathcal{E}(\mathcal{D}|M)$ represents the evidence of a model $M$ considering the data $\mathcal{D}$. Accord, a positive (negative) value of $2\ln{\mathcal{B}}$ signifies an inclination towards IDE ($\Lambda$CDM).

The cosmological parameters in our analysis are:
\begin{equation}
    \{\Omega_\text{b} h^2, \Omega_\text{c} h^2, 100\theta_\text{s}, \ln(10^{10} A_\text{s}), n_\text{s}, \tau, w, \lambda_{1(2)}\}\,.
\end{equation}
The influence of curvature, neutrino masses, and the effective number of neutrino species is beyond the scope of this work. We therefore fix these parameters to their standard values: $\Omega_K = 0$, $\Sigma m_\nu = 0.06$ eV, and $N_{\rm eff} = 3.046$. We adopt flat priors, as listed in Table~\ref{tab:priors}. During internal tests, we observed that the nested sampling algorithm occasionally converged to a subdominant local maximum of the posterior when the prior ranges were too broad. This behaviour is expected when combining datasets that exhibit some degree of tension, such as Planck and SH0ES, which can lead to a multimodal posterior distribution. If the algorithm is not configured to handle multimodality, or if the number of live points is insufficient, it may become stuck in a local mode. To deal with this issue without significantly increasing computational cost, we explored different prior ranges to locate the dominant maximum of the posterior more efficiently. We used the $\chi^2$ value as a diagnostic to identify the correct solution. The final results, including best-fit values and 68\% credible intervals for the parameters, are presented in Tables~\ref{tab1}, \ref{tab2}, and~\ref{tab3}. For all MCMC runs shown here, we use 250 live points and set the convergence criterion to a likelihood ratio of $10^{-4}$ between the live points and the final sample.

\begin{table}[htbp]
\caption{Priors}
\label{tab:priors}
\centering
\setlength\tabcolsep{2.2pt}
\renewcommand{\arraystretch}{1.1} 
\begin{tabular}{ccccc}
 \toprule
 \toprule
Parameter & \multicolumn{4}{c}{Priors} \\
\midrule
$\Omega_\text{b} h^2$ & \multicolumn{4}{c}{$[0.020,0.024]$} \\
$\Omega_\text{c} h^2$ & \multicolumn{4}{c}{$[0.00,0.14]$} \\
$100\theta_\text{s}$ & \multicolumn{4}{c}{$[1.03,1.05]$} \\
$\ln(10^{10} A_\text{s})$ & \multicolumn{4}{c}{$[2.90,3.20]$} \\
$n_\text{s}$ & \multicolumn{4}{c}{$[0.92,1.00]$} \\
$\tau$ & \multicolumn{4}{c}{$[0.01,0.12]$} \\
\midrule
 & Model I & Model II & Model III & Model IV\\
 \midrule
$w$ & $[-3.0,-0.3]$ & $[-3.0,-1.0]$ & $[-3.0,-1.0]$ & $-0.999$ \\
$\lambda_{1(2)}$ & $[-1.5,1.5]$ & $[0.0,0.04]$ & $[0.0,0.04]$ & $[-1.5,0.0]$ \\
\bottomrule
\bottomrule
\end{tabular}
\end{table}

\begin{figure*}
    \centering
    \includegraphics[width=\textwidth]{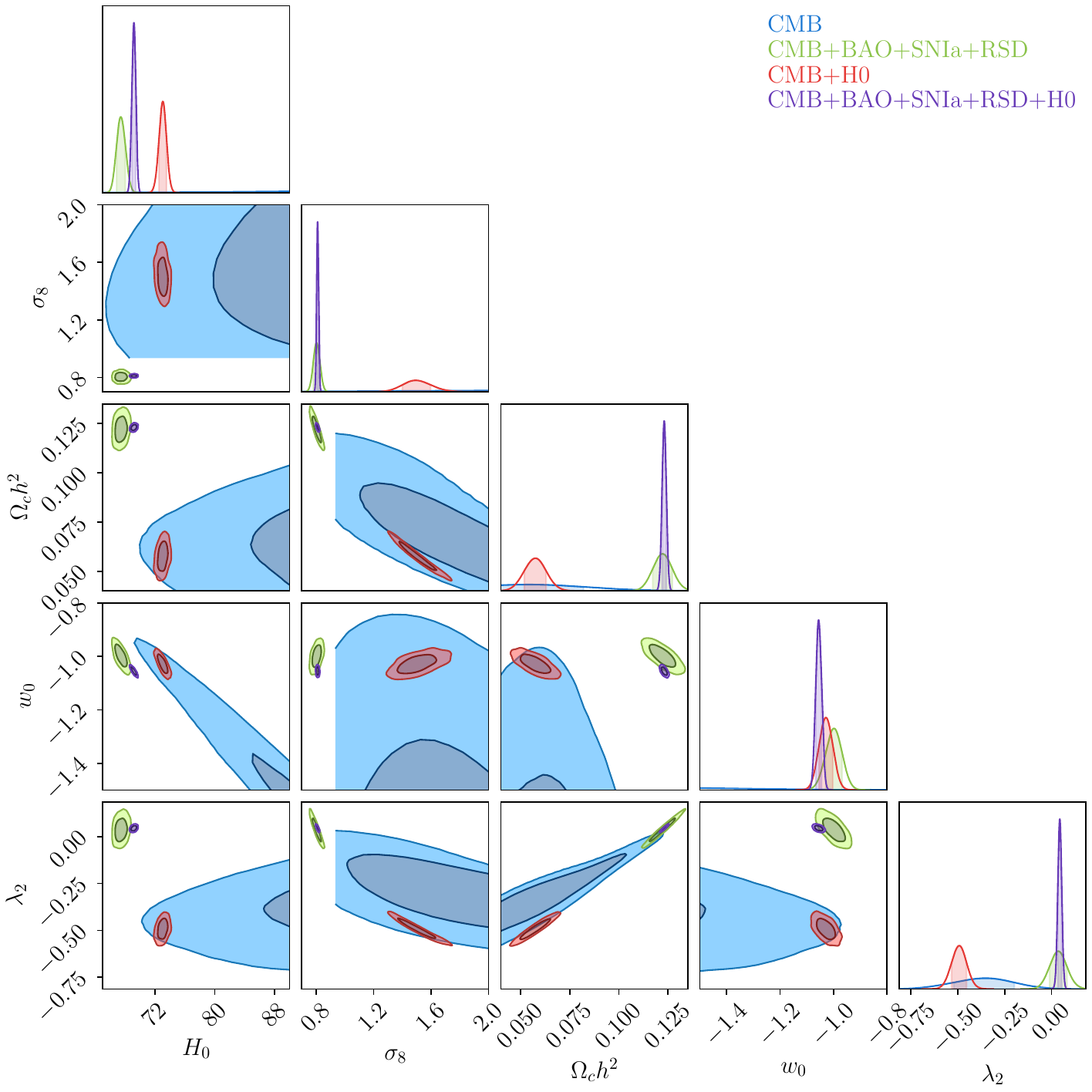}
    \caption{2-D contours (68\% and 95\% C.L.) of selected cosmological parameters for Model I, using different combinations of datasets. The inclusion of datasets other than CMB improves significantly the constraints. The different contours, however, do not superimpose and the degeneracy between $\Omega_c h^2$ and $\lambda_2$ and between $w$ and $H_0$ are not broken by the combination of data.  }
    \label{TP_model1}
\end{figure*}

\begin{figure*}
    \centering
    \includegraphics[width=\textwidth]{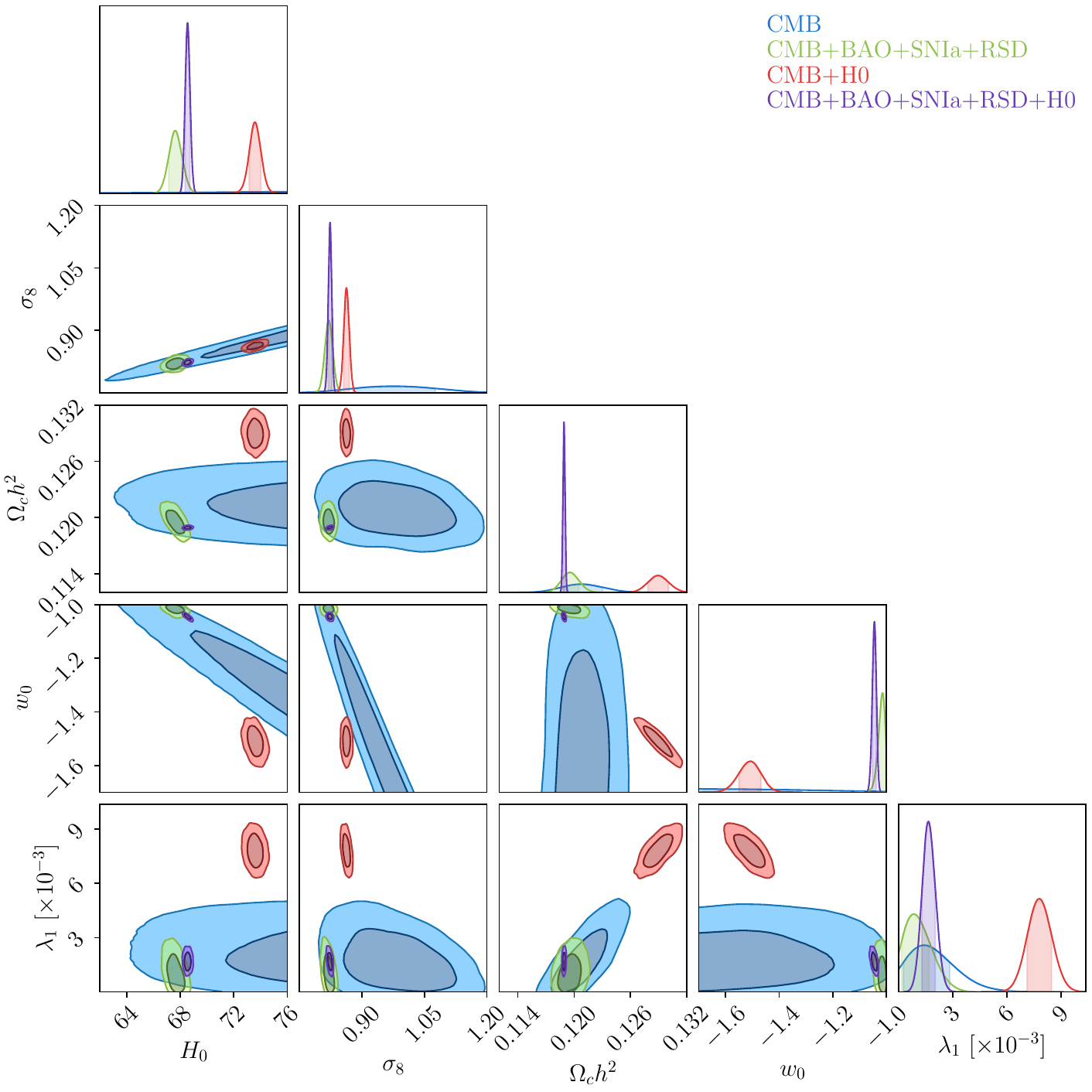}
    \caption{2-D contours (68\% and 95\% C.L.) of selected cosmological parameters for Model II, using different combinations of datasets. Contrary to Model I, the degeneracies between some cosmological parameters are broken by the combination of datasets. The contours of all datasets (but SH0ES) are overlapped with only CMB. When SH0ES is taken into account with CMB the contours no longer superimpose, showing that even if the Hubble tension is alleviated it cannot be solved by this model. }
    \label{TP_model2}
\end{figure*}

\begin{figure*}
    \centering
    \includegraphics[width=\textwidth]{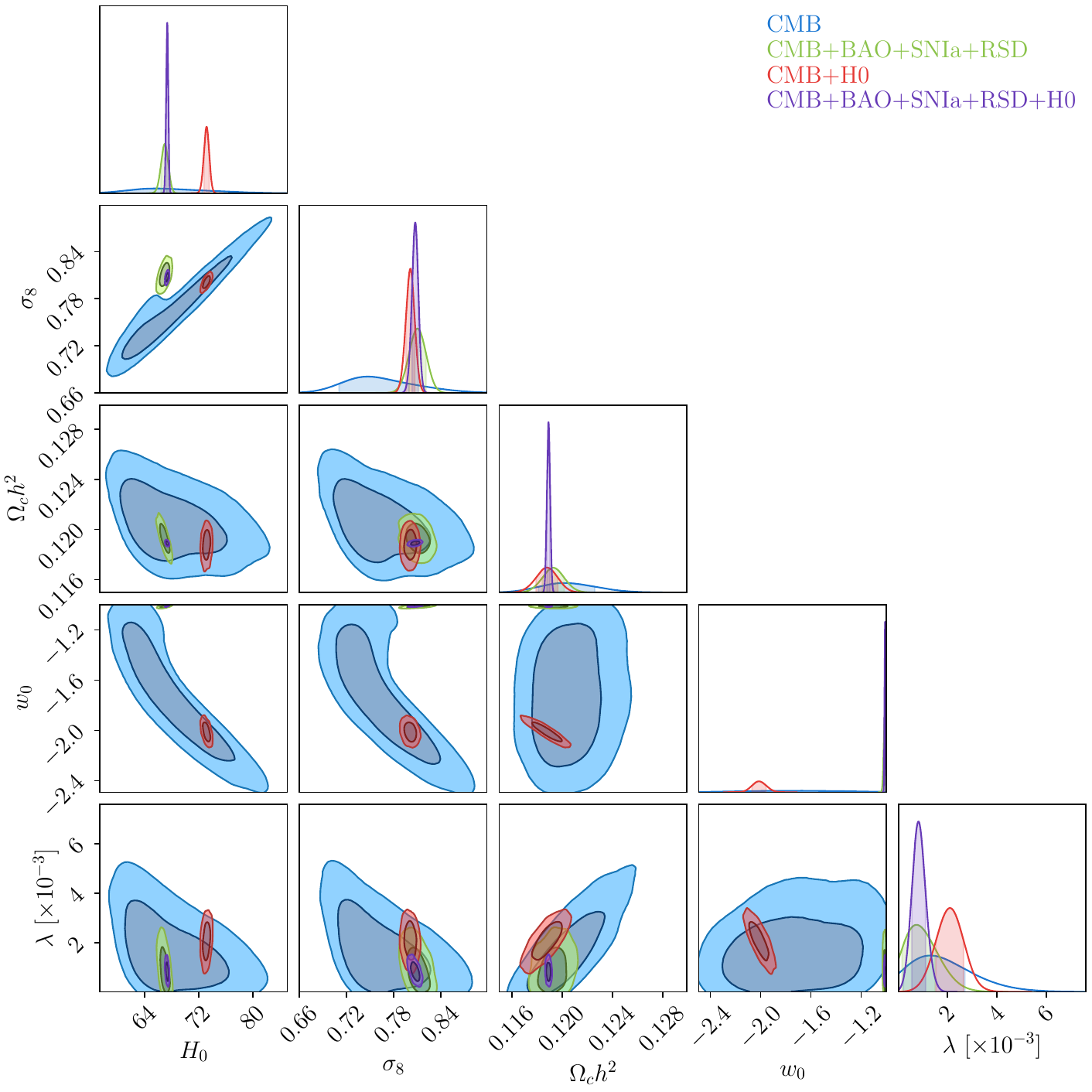}
    \caption{2-D contours (68\% and 95\% C.L.) of selected cosmological parameters for Model III, using different combinations of datasets. Similarly to Model II, the degeneracies are broken by the combination of datasets and the contours overlap for some parameters, although not in all the cases.  }
    \label{TP_model3}
\end{figure*}

In contrast to $\Lambda$CDM, some cosmological parameters in the models remain poorly constrained when using only CMB data, as also noted in \cite{Costa:2016tpb}. In particular, $\Omega_c h^2$ and $H_0$ have values very different from the ones in $\Lambda$CDM. However, from Figures \ref{TP_model1}, \ref{TP_model2} and \ref{TP_model3} we see that $w$ and $H_0$ degenerate while there is also a degeneration between $\Omega_c h^2$ and the coupling constant, indicating that $\Lambda$CDM can be recovered and a deviation in one parameter brings the other one away from the $\Lambda$CDM value. The combination of CMB with BAO, RSD, and SNIa reduces the uncertainties on the parameters, however, the resulting contours do not overlap for Model I. The low-redshift data suggest a preference for values of the parameters that are consistent with the ones of $\Lambda$CDM. The degeneracies aforementioned are maintained for Model I, although with more constrained contours, while for Models II and III they are broken in some cases. 

The non-overlapping of the joint contours of all datasets with the CMB one for Model I suggests that this model is not compatible with all cosmological data. A similar situation happens for Model III for the $w-H_0$ contour.  Only Model II is consistent with all datasets (but $H_0$).

The value of $H_0$ using CMB data alone is large enough to possibly alleviate the Hubble tension and to investigate its impact on the results we use the SH0ES Cepheid host distance anchors in the likelihood as well, producing two joint analyses with CMB. The parameters are then even more constrained, however, the contours are no longer overlapped for all data, as they were in some cases.

When all datasets are considered together $H_0$ has its discrepancy reduced to $0.13\sigma$ ($5.4\sigma$) when compared to the standard $\Lambda$CDM (R22) value for Model III, while for Models I and II the values are $3.2\sigma$ and $2.2\sigma$ ($3.6\sigma$ and $4.3\sigma$) away from 
\textit{Planck} (R22), respectively.  However, only Model II presents a consistent overlap of CMB and  CMB+BAO+SNIa+RSD contours. When all datasets are analyzed together, only for some parameters the 2-D contours are overlapped, indicating that  SH0ES is not compatible with BAO, RSD, and SNIa for any IDE model assumed here.

\begin{figure*}
    \centering
    \includegraphics[width=\textwidth]{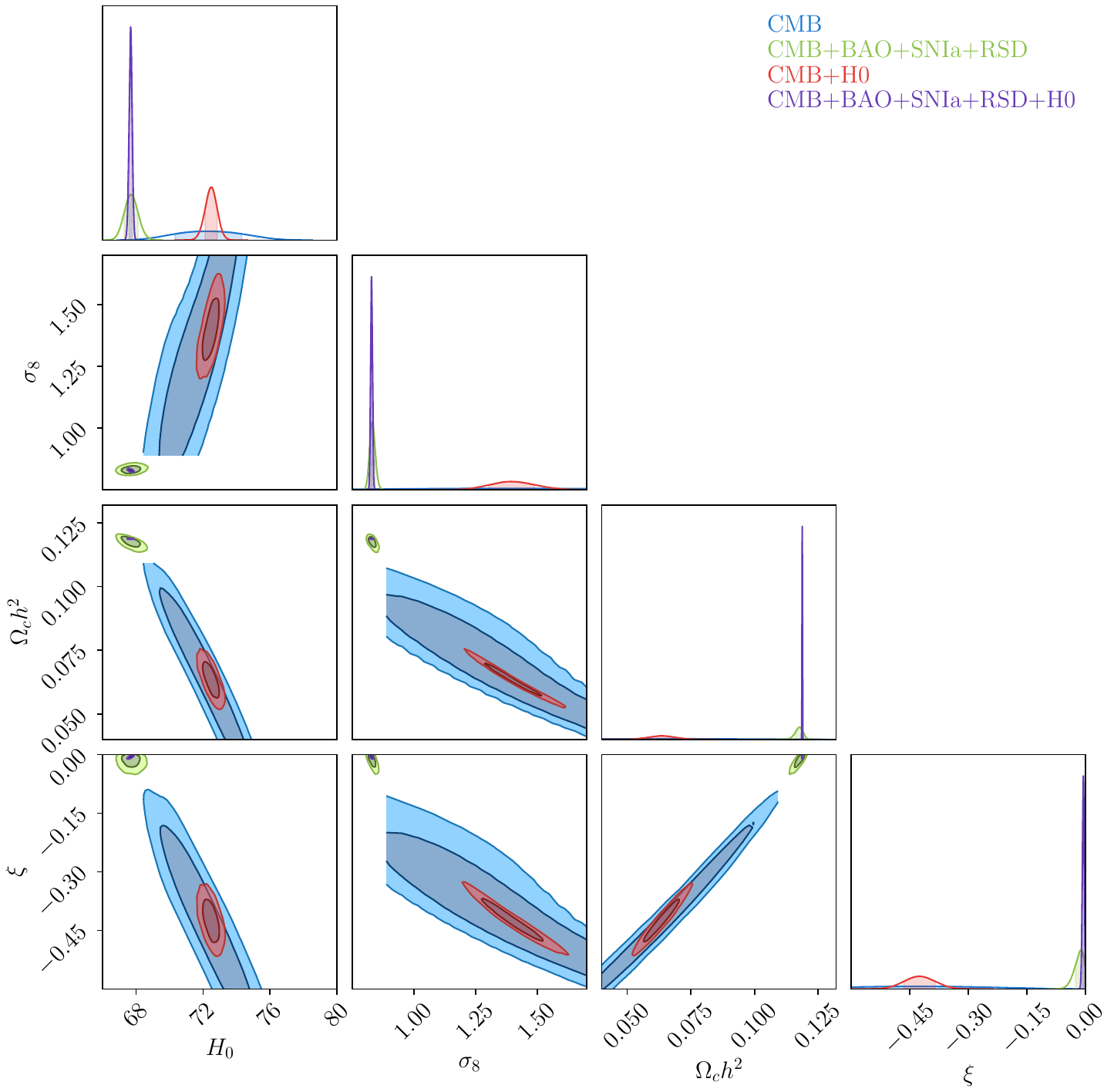}
    \caption{2-D contours (68\% and 95\% C.L.) of selected cosmological parameters for Model IV, using different combinations of datasets. Here the degeneracies are not broken by the inclusion of other datasets and the contours mostly do not overlap. The shape of the contours is very similar to what was obtained in previous work \cite{Lucca:2020zjb}.}
    \label{TP_model4}
\end{figure*}

Moreover, we constrained the model presented in Sec. \ref{sec:lucca}, whose results are found in Table \ref{tab4} and Fig. \ref{TP_model4}. They reproduced partially the findings in \cite{Lucca:2020zjb}, but the difference here is the inclusion of more recent datasets (such as Pantheon+) and the usage of R22 instead of \cite{Riess:2019cxk}. The results for the cosmological parameters only using \textit{Planck} are equivalent, however, the other combinations of surveys produce generally tighter constraints.  The inclusion of all datasets but $H_0$ produces an incompatibility with the case of CMB alone. Here we used more cosmological data than \cite{Lucca:2020zjb}. The value of $H_0$, for the combination of all datasets, gives a discrepancy of $0.6\sigma$ from $\Lambda$CDM (instead of the previously found $2.5\sigma$) and $5.1\sigma$ from R22 (instead of $2.6\sigma$). The Hubble tension is therefore not alleviated, because the value of $H_0$ is very similar to the $\Lambda$CDM one. 
The joint contours of all datasets are roughly consistent with CMB alone or CMB + $H_0$.

Regarding the values of $\sigma_8$ inferred here, they are between 0.8 and 1.9 with the exception of Model III which reaches 0.75 (CMB), although its best-fit is 0.84. The 2-D plots are not completely overlapped when we assume the different datasets.  The results for $\sigma_8$ do not seem to help much to alleviate the $\sigma_8$ tension, as the values are close to the \textit{Planck} one (see more details about this tension in \cite{DiValentino:2020vvd}).

Finally, we use the Bayes ratio in Eq.(\ref{bayes}) to compute the values of $2\ln{\mathcal{B}}$ shown in Table \ref{stats}. The results indicate a significant preference for IDE models when considering individual datasets such as CMB, CMB+BAO+SNIa+RSD, and CMB+$H_0$. When combining all datasets, the preference shifts towards $\Lambda$CDM in all cases except for Model I. However, we cannot conclude that there is strong evidence in favour of Model I due to the inconsistency among the joint contours from various datasets. This tension may bias the Bayesian evidence, making any comparison between models also biased.

\begin{table*}[htbp]
\caption{$\Delta \chi ^2 = \chi_{\text{IDE}}^2 - \chi_{\Lambda\text{CDM}}^2$ and evidence level in $2 \ln{\mathcal{B}}$ scale compared to $\Lambda$CDM. The $\chi^2$ value for each model was calculated as $\chi^2 = -2\ln{\mathcal{L}}$, where $\mathcal{L}$ is the maximum likelihood.}
\label{stats}
\centering
\setlength\tabcolsep{12.5pt} 
\renewcommand{\arraystretch}{1.2} 
\begin{tabular}{c cc cc cc cc}
 \toprule
 \toprule
 \multirow{2}{*}{Model} & \multicolumn{2}{c}{CMB} & \multicolumn{2}{c}{CMB+BAO+SNIa+RSD} & \multicolumn{2}{c}{CMB+H$_0$} & \multicolumn{2}{c}{CMB+BAO+SNIa+RSD+H$_0$} \\
 \cmidrule(lr){2-9}
   & $\Delta \chi ^2$ & $2\ln \mathcal{B}$ & \hspace{.5cm}  $\Delta \chi ^2$ & \hspace{-.5cm} $2\ln \mathcal{B}$ & $\Delta \chi ^2$ & $2\ln \mathcal{B}$ &\hspace{.8cm}  $\Delta \chi ^2$ &\hspace{-.8cm}  $2\ln \mathcal{B}$ \\
 \midrule
 I   & $-3.7$  & $4.8$  & \hspace{.5cm} $-1.2$  & \hspace{-.5cm} $15.4$ & $-20.5$ & $19.3$ &\hspace{.8cm}  $-25.2$ &\hspace{-.8cm}  $8.5$ \\
 II  & $-2.6$  & $2.5$  & \hspace{.5cm} $-0.2$  & \hspace{-.5cm} $15.3$ & $-14.7$ & $15.7$ &\hspace{.8cm}  $-13.5$ &\hspace{-.8cm}  $-0.8$ \\
 III & $-2.5$  & $2.5$  & \hspace{.5cm} $-1.1$  & \hspace{-.5cm} $15.9$ & $-18.8$ & $17.1$ &\hspace{.8cm}  $-5.2$  &\hspace{-.8cm}  $-11.0$ \\
 IV  & $-0.1$  & $2.3$  & \hspace{.5cm} $-0.5$  & \hspace{-.5cm} $15.2$ & $-25.6$ & $27.3$ &\hspace{.8cm}  $-6.2$  &\hspace{-.8cm}  $-9.8$ \\
\bottomrule
\bottomrule
\end{tabular}
\end{table*}


\begin{table*}[htbp]
\caption{Cosmological parameters -- Model I}
\label{tab1}
\centering
\setlength\tabcolsep{3.1pt}
\renewcommand{\arraystretch}{1.2} 
\begin{tabular}{ccccccccc}
 \toprule
 \toprule
 \multirow{2}{*}{Parameter } & \multicolumn{2}{c}{CMB} & \multicolumn{2}{c}{CMB+BAO+SNIa+RSD} & \multicolumn{2}{c}{CMB+H$_0$} & \multicolumn{2}{c}{CMB+BAO+SNIa+RSD+H$_0$} \\
 \cmidrule{2-9}
   & Best fit & 68\% limits & Best fit & 68\% limits & Best fit & 68\% limits & Best fit & 68\% limits \\
   
 \midrule
 
 $\Omega_\text{b} h^2$ & $0.02229$ & $0.02243 \pm 0.00016$ & $0.02247$ & $0.02244 ^{+0.00015}_{-0.00014}$ & $0.02240$ & $0.02232^{+0.000085}_{-0.000087}$ & $0.022445$ & $0.022431 ^{+0.000046}_{-0.000047}$ \\
 
 $\Omega_\text{c} h^2$ & $0.124$ & $0.054^{+0.028}_{-0.029}$ & $0.1228$ & $0.1222 \pm 0.005$ & $0.0607$ & $0.0576^{+0.0055}_{-0.0057}$ & $0.1232$ & $0.1230 \pm 0.0011$ \\
 
 $100\theta_\text{S}$ & $1.0419$ & $1.0420 \pm 0.00029$ & $1.04199$ & $1.04201^{+0.00029}_{-0.00028}$ & $1.04193$ & $1.04187^{+0.00016}_{-0.00017}$ & $1.04206$ & $1.04209 ^{+0.00015}_{-0.00014}$ \\
 
 $\ln(10^{10} A_\text{s})$ & $3.018$ & $3.039\pm 0.027$ & $3.095$ & $3.092^{+0.025}_{-0.024}$ & $3.067$ & $3.061\pm 0.014$ & $3.0763$ & $3.0830^{+0.0081}_{-0.0076}$ \\
 
 $n_\text{s}$ & $0.9628$ & $0.9682 \pm 0.0044$ & $0.9676$ & $0.9681^{+0.0040}_{-0.0037}$ & $0.9670$ & $0.9658\pm 0.0024$ & $0.9685$ & $0.9695 ^{+0.0015}_{-0.0014}$ \\

 $\tau $ & $0.0541$ & $0.0501^{+0.0071}_{-0.0078}$ & $0.0526$ & $0.0517^{+0.0075}_{-0.0074}$ & $0.0525$ & $0.0523^{+0.0031}_{-0.0032}$ & $0.0524$ & $0.0526^{+0.0031}_{-0.0032}$ \\
 
 $w_0$ & $-2.59$ & $-2.08\pm 0.47$ & $-0.997$ & $-0.998^{+0.030}_{-0.031}$ & $-1.029$ & $-1.028^{+0.025}_{-0.026}$ & $-1.054$ & $-1.056\pm{0.011}$ \\
 
 $\lambda_2$ & $0.022$ & $-0.35^{+0.15}_{-0.17}$ & $0.041$ & $0.036^{+0.046}_{-0.047}$ & $-0.460$ & $-0.492^{+0.040}_{-0.041}$ & $0.045$ & $0.044^{+0.010}_{-0.011}$ \\
 
 \midrule
 
 $H_0$ & $126$ & $110^{+20}_{-18}$ & $67.34$ & $67.43^{+0.60}_{-0.58}$ & $73.15$ & $73.05^{+0.49}_{-0.52}$ & $69.10$ & $69.18^{+0.26}_{-0.27}$ \\
 
 $\sigma_8$ & $1.19$ & $1.93^{+1.01}_{-0.82}$ & $0.801$ & $0.803\pm0.023$ & $1.434$ & $1.492^{+0.105}_{-0.093}$ & $0.8067$ & $0.8103^{+0.0068}_{-0.0065}$ \\
 
 $Age/Gyr$ & $13.339$ & $13.367^{+0.102}_{-0.094}$ & $13.800$ & $13.800\pm 0.020$ & $13.699$ & $13.709\pm 0.013$ & $13.7649$ & $13.7635^{+0.0078}_{-0.0071}$ \\
 
$S_8$ & $0.66$ & $0.91^{+0.21}_{-0.17}$ & $0.827$ & $0.827\pm0.014$ & $1.032$ & $1.056^{+0.038}_{-0.033}$ & $0.8135$ & $0.8162^{+0.0053}_{-0.0058}$ \\ 

\midrule
$\chi^2_{\text{min}}$ & \multicolumn{2}{c}{$2762.45$} & \multicolumn{2}{c}{$4206.31$} & \multicolumn{2}{c}{$6339.80$} & \multicolumn{2}{c}{$7784.30$} \\
\bottomrule
\bottomrule
\end{tabular}
\end{table*}

\begin{table*}[htbp]
\caption{Cosmological parameters -- Model II}
\label{tab2}
\centering
\setlength\tabcolsep{1.7pt}
\renewcommand{\arraystretch}{1.2} 
\begin{tabular}{ccccccccc}
 \toprule
 \toprule
 \multirow{2}{*}{Parameter } & \multicolumn{2}{c}{CMB} & \multicolumn{2}{c}{CMB+BAO+SNIa+RSD} & \multicolumn{2}{c}{CMB+H$_0$} & \multicolumn{2}{c}{CMB+BAO+SNIa+RSD+H$_0$} \\
 \cmidrule{2-9}
   & Best fit & 68\% limits & Best fit & 68\% limits & Best fit & 68\% limits & Best fit & 68\% limits \\
   
 \midrule
 
 $\Omega_\text{b} h^2$ & $0.02237$ & $0.02248\pm 0.00017$ & $0.02238$ & $0.02246 \pm 0.00015$ & $0.022481$ & $0.022514^{+0.000069}_{-0.000075}$ & $0.02251$ & $0.02251\pm0.000044$ \\
 
 $\Omega_\text{c} h^2$ & $0.1201$ & $0.1207^{+0.0023}_{-0.0022}$ & $0.11878$ & $0.11956^{+0.00095}_{-0.00094}$ & $0.1284$ & $0.1290\pm 0.0011$ & $0.11885$ & $0.11891\pm0.00011$ \\
 
 $100\theta_\text{S}$ & $1.04191$ & $1.04194^{+0.00031}_{-0.00030}$ & $1.04201$ & $1.04197 ^{+0.00029}_{-0.00030}$ & $1.04164$ & $1.04164\pm 0.00014$ & $1.04203$ & $1.04202\pm0.00015$ \\
 
 $\ln(10^{10} A_\text{s})$ & $3.020$ & $3.045\pm 0.028$ & $3.077$ & $3.081\pm0.024$ & $3.031$ & $3.030\pm 0.013$ & $3.0849$ & $3.0884\pm0.0077$ \\
 
 $n_\text{s}$ & $0.9654$ & $0.9654^{+0.0050}_{-0.0045}$ & $0.9661$ & $0.9670^{+0.0038}_{-0.0037}$ & $0.9585$ & $0.9575\pm 0.0023$ & $0.9672$ & $0.9672\pm0.00011$ \\
 
 $\tau $ & $0.0483$ & $0.0497^{+0.0078}_{-0.0076}$ & $0.0516$ & $0.0510^{+0.0072}_{-0.0077}$ & $0.0508$ & $0.0503^{+0.0037}_{-0.0035}$ & $0.0509$ & $0.0518^{+0.0027}_{-0.0025}$ \\
 
 $w_0$ & $-2.19$ & $-1.70^{+0.38}_{-0.37}$ & $-1.006$ & $-1.014^{+0.011}_{-0.014}$ & $-1.484$ & $-1.507^{+0.039}_{-0.043}$ & $-1.0390$ & $-1.0442^{+0.0074}_{-0.0076}$ \\
 
 $\lambda_1$ & $0.00068$ & $0.0014^{+0.0014}_{-0.0011}$ & $0.000012$ & $0.00086^{+0.00085}_{-0.00068}$ & $0.00742$ & $0.00778^{+0.00068}_{-0.00065}$ & $0.00133$ & $0.00167^{+0.00038}_{-0.00036}$ \\
 
 \midrule
 
 $H_0$ & $108$ & $87^{+15}_{-13}$ & $67.89$ & $67.63\pm0.49$ & $73.50$ & $73.57^{+0.46}_{-0.41}$ & $68.55$ & $68.55 \pm 0.18$ \\
 
 $\sigma_8$ & $1.110$ & $0.975^{+0.101}_{-0.096}$ & $0.8224$ & $0.8204^{+0.0094}_{-0.0098}$ & $0.8624$ & $0.8627^{+0.0067}_{-0.0065}$ & $0.8232$ & $0.8233^{+0.0040}_{-0.0041}$ \\
 
 $Age/Gyr$ & $13.42$ & $13.56^{+0.15}_{-0.13}$ & $13.795$ & $13.817^{+0.027}_{-0.026}$ & $13.902$ & $13.910\pm 0.022$ & $13.802$ & $13.812^{+0.011}_{-0.010}$ \\
 
 $S_8$ & $0.707$ & $0.769^{+0.042}_{-0.045}$ & $0.831$ & $0.835\pm0.012$ & $0.8320$ & $0.8330\pm0.0063$ & $0.8243$ & $0.8245^{+0.0039}_{-0.0040}$ \\ 
 
\midrule
$\chi^2_{\text{min}}$ & \multicolumn{2}{c}{$2763.53$} & \multicolumn{2}{c}{$4207.37$} & \multicolumn{2}{c}{$6345.58$} & \multicolumn{2}{c}{$7796.09$} \\
\bottomrule
\bottomrule
\end{tabular}
\end{table*}

\begin{table*}[htbp]
\caption{Cosmological parameters -- Model III}
\label{tab3}
\centering
\setlength\tabcolsep{2.3pt}
\renewcommand{\arraystretch}{1.2} 
\begin{tabular}{ccccccccc}
 \toprule
 \toprule
 \multirow{2}{*}{Parameter} & \multicolumn{2}{c}{CMB} & \multicolumn{2}{c}{CMB+BAO+SNIa+RSD} & \multicolumn{2}{c}{CMB+H$_0$} & \multicolumn{2}{c}{CMB+BAO+SNIa+RSD+H$_0$} \\
 \cmidrule{2-9}
   & Best fit & 68\% limits & Best fit & 68\% limits & Best fit & 68\% limits & Best fit & 68\% limits \\
   
 \midrule
 
 $\Omega_\text{b} h^2$ & $0.02237$ & $0.02249\pm 0.00016$ & $0.02233$ & $0.02246^{+0.00015}_{-0.00014}$ & $0.022603$ & $0.022645^{+0.000078}_{-0.000082}$ & $0.022578$ & $0.022575\pm0.000034$ \\
 
 $\Omega_\text{c} h^2$ & $0.1199$ & $0.1202^{+0.0024}_{-0.0021}$ & $0.11896$ & $0.11928^{+0.00088}_{-0.00087}$ & $0.11914$ & $0.11877^{+0.00088}_{-0.00086}$ & $0.11891$ & $0.11891^{+0.00012}_{-0.00013}$ \\
 
 $100\theta_\text{S}$ & $1.04201$ & $1.04197^{+0.00029}_{-0.00030}$ & $1.04198$ & $1.04202^\pm0.00028$ & $1.04217$ & $1.04220\pm 0.00016$ & $1.04215$ & $1.04217\pm0.00013$ \\
 
 $\ln(10^{10} A_\text{s})$ & $3.005$ & $3.017\pm 0.027$ & $3.063$ & $3.082^{+0.024}_{-0.023}$ & $3.040$ & $3.049^{+0.011}_{-0.012}$ & $3.0774$ & $3.0758^{+0.0072}_{-0.0074}$ \\
 
 $n_\text{s}$ & $0.9664$ & $0.9670^{+0.0046}_{-0.0047}$ & $0.9653$ & $0.9676 \pm 0.0037$ & $0.9692$ & $0.9707\pm 0.0022$ & $0.96559$ & $0.96575^{+0.00104}_{-0.0099}$ \\
 
 $\tau$ & $0.0519$ & $0.0493\pm 0.0078$ & $0.0507$ & $0.0514^{+0.0071}_{-0.0075}$ & $0.0476$ & $0.0483\pm 0.0028$ & $0.0510$ & $0.0510\pm0.0029$ \\
 
 $w_0$ & $-2.26$ & $-1.77^{+0.44}_{-0.36}$ & $-1.0100$ & $-1.0114^{+0.0073}_{-0.0082}$ & $-2.036$ & $-2.014^{+0.057}_{-0.055}$ & $-1.0078$ & $-1.0099^{+0.0035}_{-0.0038}$ \\
 
 $\lambda$ & $0.00057$ & $0.0013^{+0.0014}_{-0.0011}$ & $0.00020$ & $0.00074^{+0.00074}_{-0.00059}$ & $0.00202$ & $0.00210^{+0.00055}_{-0.00057}$ & $0.00072$ & $0.00082^{+0.00029}_{-0.00027}$ \\
 
 \midrule
 
 $H_0$ & $77.5$ & $65.6^{+6.9}_{-4.4}$ & $67.20$ & $66.96^{+0.52}_{-0.55}$ & $73.26$ & $73.15^{+0.39}_{-0.41}$ & $67.43$ & $67.33^{+0.15}_{-0.16}$ \\
 
 $\sigma_8$ & $0.843$ & $0.747^{+0.053}_{-0.037}$ & $0.809$ & $0.810\pm0.011$ & $0.8016$ & $0.8013^{+0.0056}_{-0.0058}$ & $0.8100$ & $0.8075^{+0.0040}_{-0.0042}$ \\
 
 $Age/Gyr$ & $13.40$ & $13.53^{+0.16}_{-0.13}$ & $13.807$ & $13.812^{+0.026}_{-0.024}$ & $13.453$ & $13.450\pm 0.018$ & $13.792$ & $13.793^{+0.0100}_{-0.0092}$ \\
 
  $S_8$ & $0.749$ & $0.772^{+0.027}_{-0.026}$ & $0.826$ & $0.832^{+0.011}_{-0.010}$ & $0.7521$ & $0.7520^{+0.0044}_{-0.0043}$ & $0.8250$ & $0.8235\pm{+0.0041}$ \\
  
\midrule
$\chi^2_{\text{min}}$ & \multicolumn{2}{c}{$2763.66$} & \multicolumn{2}{c}{$4206.40$} & \multicolumn{2}{c}{$6341.54$} & \multicolumn{2}{c}{$7804.30$} \\
\bottomrule
\bottomrule
\end{tabular}
\end{table*}

\begin{table*}[htbp]
\caption{Cosmological parameters -- Model IV}
\label{tab4}
\centering
\setlength\tabcolsep{1.7pt}
\renewcommand{\arraystretch}{1.2} 
\begin{tabular}{ccccccccc}
 \toprule
 \toprule
 \multirow{2}{*}{Parameter } & \multicolumn{2}{c}{CMB} & \multicolumn{2}{c}{CMB+BAO+SNIa+RSD} & \multicolumn{2}{c}{CMB+H$_0$} & \multicolumn{2}{c}{CMB+BAO+SNIa+RSD+H$_0$} \\
 \cmidrule{2-9}
   & Best fit & 68\% limits & Best fit & 68\% limits & Best fit & 68\% limits & Best fit & 68\% limits \\
   
 \midrule
 
 $\Omega_\text{b} h^2$ & $0.02244$ & $0.02241\pm 0.00016$ & $0.02238$ & $0.02239 \pm 0.00014$ & $0.022436$ & $0.022443^{+0.000077}_{-0.000076}$ & $0.022451$ & $0.022451 \pm 0.000037$ \\
 
 $\Omega_\text{c} h^2$ & $0.095$ & $0.068^{+0.023}_{-0.031}$ & $0.1191$ & $0.1177^{+0.0015}_{-0.0016}$ & $0.0656$ & $0.0634^{+0.0052}_{-0.0051}$ & $0.118718$ & $0.118731^{+0.000091}_{-0.000087}$ \\
 
 $100\theta_\text{S}$ &$1.04196$ & $1.04199^{+0.00029}_{-0.00030}$ & $1.04196$ & $1.04196\pm0.00029$ & $1.04199$ & $1.04198^{+0.00014}_{-0.00015}$ & $1.04219$ & $1.04222^{+0.00013}_{-0.00012}$ \\
 
 $\ln(10^{10} A_\text{s})$ & $3.065$ & $3.069\pm 0.028$ & $3.072$ & $3.077\pm0.024$ & $3.082$ & $3.084\pm 0.014$ & $3.0810$ & $3.0806^{+0.0069}_{-0.0070}$ \\
 
 $n_\text{s}$ & $0.9656$ & $0.9682\pm 0.0042$ & $0.9654$ & $0.9666^{+0.0037}_{-0.0038}$ & $0.9699$ & $0.9706^{+0.0023}_{-0.0020}$ & $0.96914$ & $0.96885^{+0.00091}_{-0.00099}$ \\
 
 $\tau$ & $0.0496$ & $0.0496\pm 0.0079$ & $0.0503$ & $0.0505^{+0.0076}_{-0.0075}$ & $0.0557$ & $0.0555^{+0.0039}_{-0.0040}$ & $0.0519$ & $0.0518 \pm 0.0028$ \\
 
 $\xi$ & $-0.21$ & $-0.42^{+0.17}_{-0.20}$ & $-0.00031$ & $-0.012^{+0.010}_{-0.014}$ & $-0.413$ & $-0.426^{+0.041}_{-0.039}$ & $-0.0044$ & $-0.0057\pm 0.0024$ \\
 
 \midrule
 
 $H_0$ & $69.8$ & $72.2^{+2.1}_{-1.9}$ & $67.53$ & $67.70\pm0.42$ & $72.27$ & $72.49^{+0.35}_{-0.36}$ & $67.702$ & $67.675\pm0.090$ \\
 
 $\sigma_8$ & $1.00$ & $1.33^{+0.64}_{-0.38}$ & $0.820$ & $0.832^{+0.012}_{-0.011}$ & $1.364$ & $1.394^{+0.095}_{-0.089}$ & $0.8273$ & $0.8283^{+0.0037}_{-0.0035}$ \\
 
 $Age/Gyr$ & $13.752$ & $13.710^{+0.040}_{-0.039}$ & $13.803$ & $13.798^{+0.020}_{-0.019}$ & $13.710$ & $13.707^{+0.011}_{-0.012}$ & $13.7878$ & $13.7866^{+0.0061}_{-0.0056}$ \\
 
 $S_8$ & $0.90$ & $1.01^{+0.20}_{-0.15}$ & $0.834$ & $0.839\pm0.011$& $1.022$ & $1.031^{+0.035}_{-0.033}$ & $0.8382$ & $0.8396^{+0.0040}_{-0.0043}$\\ 
 
\midrule
$\chi^2_{\text{min}}$ & \multicolumn{2}{c}{$2766.04$} & \multicolumn{2}{c}{$4207.00$} & \multicolumn{2}{c}{$6334.64$} & \multicolumn{2}{c}{$7803.40$} \\

\bottomrule
\bottomrule
\end{tabular}
\end{table*}

\section{Conclusions}\label{sec:conclusions}

Interacting Dark Energy has been an alternative to the standard $\Lambda$CDM model for many years. With the release of new datasets, the free parameters of the model can be even more constrained, reaching a level of precision not obtained before.

In this paper, we have investigated a class of coupled DE models, where
DE is interacting with DM phenomenologically. We have included a term in the perturbation equations for both fluids that was absent in previous works but is required by gauge invariance and considers the perturbation of the Hubble rate. Furthermore, we have obtained updated constraints on the cosmological parameters, using more recent data from CMB anisotropies, BAO, RSD, and SNIa, showing tighter constraints on the coupling constants.  However, Model III and especially Model I show incompatibility with every cosmological data because we can see that the contours do not overlap when different datasets constrain the cosmological parameters.  

We used the SH0ES Cepheid host distance anchors combined with other likelihoods to constrain $H_0$ in the different models. Model I is the one that presents a better alleviation of the Hubble tension. Model IV also produces a lower discrepancy of  $H_0$, although the new data (including the new result from SH0ES, R22) made the tension less alleviated than the previous work \cite{Lucca:2020zjb}. However, when all datasets are taken into account, no overlap between the 2D contours exists for any of the models, still indicating an incompatibility between low-redshift data and $H_0$.

Finally, we performed a statistical model comparison, and a preference for IDE over $\Lambda$CDM happens for all the models, when the datasets CMB, BAO, RSD, and SNIa are included. However, the 2D plots do not overlap for the majority of models.

We conclude that the IDE models considered in this paper are not flexible enough to fit all cosmological data including values of $H_0$ from SH0ES in a statistically acceptable way. The already existing tension in $\Lambda$CDM cannot be significantly alleviated in the scenario analyzed here, therefore, the models would need to be modified to include further flexibility of predictions.


\begin{acknowledgements}
\noindent R.G.L. thanks Matteo Lucca for the useful comments. G.A.H acknowledges the funding Dean's Doctoral Scholarship by the University of Manchester and a CAPES grant 8887.622333/2021-00. L.O.P. acknowledges a CNPq grant 137172/2022-2. R.P.R. acknowledges a CNPq grant 125103/2022-0.  This work was developed in the Brazilian cluster SDumont. We thank the anonymous referee for the very valuable report. 
\end{acknowledgements}

\bibliography{main.bbl}\end{document}